\documentclass[aps,preprint]{revtex4}%
\usepackage{amsfonts}
\usepackage{amsmath}
\usepackage{amssymb}
\usepackage{graphicx}%
\begin{document}
\preprint{ }
\title{Zone Determinant Expansions for Nuclear Lattice Simulations}
\author{Dean J. Lee}
\email{djlee3@unity.ncsu.edu}
\affiliation{Department of Physics, North Carolina State University, Box 8202, Raleigh, NC 27695}
\author{Ilse C.F. Ipsen}
\email{ipsen@math.ncsu.edu}
\affiliation{Center for Research in Scientific Computation, Department of Mathematics,
North Carolina State University, Box 8205, Raleigh, NC 27695}

\begin{abstract}
We discuss simulations of finite temperature nuclear matter on the lattice.
\ We introduce a new approximation to nucleon matrix determinants that is
physically motivated by chiral effective theory. \ The method involves
breaking the lattice into spatial zones and expanding the determinant in
powers of the boundary hopping parameter.

\end{abstract}
\maketitle

\section{Introduction}

We consider quantum simulations of nuclear matter on the lattice. \ In
particular we address the problem of calculating the contribution of
nucleon/nucleon-hole loops at nonzero nucleon density. \ With the help of
auxiliary boson fields, all nucleon interactions can be written in terms of
one-body interactions in a fluctuating background. \ In the grand canonical
ensemble, the contribution of nucleon/nucleon-hole loops to the partition
function equals the determinant of the one-body interaction matrix. \ Since
the determinant of the interaction matrix for a general boson field
configuration is not positive, stochastic methods such as hybrid Monte Carlo
\cite{Scalettar:1986uy}\cite{Gottlieb:1987mq}\cite{Duane:1987de} do not give
the sign or phase of the determinant. Instead one must rely on much slower and
more memory intensive algorithms based on LU\ factorization, which decomposes
matrices in terms of a product of upper and lower triangular matrices.

The number of required operations in LU factorization for an $n\times n$
matrix scales as $n^{3}$. \ It has been shown in the literature that repeated
calculations of matrix determinants with only localized changes can be
streamlined in various ways \cite{Scalapino:1981qs}\cite{Gubernatis:1992a}.
\ However it is difficult to avoid the poor scaling inherent in the method.
\ If $V$ is the spatial volume and $\beta$ is the inverse temperature measured
in lattice units, a simulation that includes nucleon/nucleon-hole loops
requires $(V\beta)^{3}$ times more operations than the corresponding quenched
simulation without loops. \ This slowdown should not be confused with the
infamous fermion sign or phase problem \footnote{We should mention recent work
\cite{Chandrasekharan:2003wy}\ that shows in the chiral limit with static
nucleons the sign problem does not occur. \ Thus there is some indirect
relationship between the two problems.} which becomes significant at
temperatures $T\leq1$ MeV. \ The computational bottleneck we are considering
is due to the inefficiencies of the algorithm and persists at all
temperatures. \ It is this numerical challenge which sets current limits on
nuclear lattice simulations.

In this paper we introduce a new approach to approximating nucleon matrix
determinants. \ We begin with a review of the current status of nuclear matter
simulations on the lattice and look to chiral effective theory to determine
the relative importance of various interactions. \ We then introduce the
concept of spatial zones and suggest a new expansion of the nucleon
determinant in powers of the hopping parameter connecting neighboring zones.
\ Rigorous bounds on the convergence of this expansion are given as well as an
estimate of the required size of the spatial zones as a function of
temperature. \ We apply the expansion to a realistic lattice simulation of the
interactions of neutrons and neutral pions.

\section{Nuclear lattice simulations}

Recently M\"{u}ller, Koonin, Seki, and van Kolck\textit{\ }%
\cite{Muller:1999cp}, pioneered the study of quantum many body effects in
infinite nuclear matter at finite density and temperature. \ In their work
they considered only nucleon degrees of freedom and used an effective
Hamiltonian of the form%
\begin{equation}
H=K+V_{c}+V_{\sigma}\text{,}%
\end{equation}
where $K$ is the kinetic energy, $V_{c}$ is the two-body scalar potential, and
$V_{\sigma}$ is the two-body tensor potential. \ While the simulation was done
on the lattice we will write their Hamiltonian in the more familiar continuum
language. \ In the continuum $K$, $V_{c}$, and $V_{\sigma}$ take the form%
\begin{equation}
K=-\tfrac{1}{2m_{N}}\int d^{3}\vec{x}\psi_{\sigma\tau}^{\dagger}\vec{\nabla
}^{2}\psi_{\sigma\tau},
\end{equation}%
\begin{equation}
V_{c}=\tfrac{1}{2}\int d^{3}\vec{x}d^{3}\vec{x}^{\prime}\psi_{\sigma\tau
}^{\dagger}(\vec{x})\psi_{\sigma^{\prime}\tau^{\prime}}^{\dagger}(\vec
{x}^{\prime})V_{c}(\vec{x}-\vec{x}^{\prime})\psi_{\sigma^{\prime}\tau^{\prime
}}(\vec{x}^{\prime})\psi_{\sigma\tau}(\vec{x}),
\end{equation}%
\begin{equation}
V_{\sigma}=\tfrac{1}{2}\int d^{3}\vec{x}d^{3}\vec{x}^{\prime}\psi_{\xi\tau
}^{\dagger}(\vec{x})\psi_{\xi^{\prime}\tau^{\prime}}^{\dagger}(\vec{x}%
^{\prime})V_{\sigma}(\vec{x}-\vec{x}^{\prime})\vec{\sigma}_{\xi\tau
\kappa\lambda}\cdot\vec{\sigma}_{\xi^{\prime}\tau^{\prime}\kappa^{\prime
}\lambda^{\prime}}\psi_{\kappa^{\prime}\lambda^{\prime}}(\vec{x}^{\prime}%
)\psi_{\kappa\lambda}(\vec{x}).
\end{equation}
In our notation summations are implied over repeated indices. $\ m_{N}$ is the
nucleon mass. $\ \psi_{\sigma\tau}^{\dagger}(\psi_{\sigma\tau})$
creates(annihilates) a nucleon of spin $\sigma$ and isospin $\tau,$ and
$\vec{\sigma}_{\xi\tau\kappa\lambda}$ are the elements of a generalized Pauli
spin-isospin matrix. \ Both potentials are assumed to have Skyrme-like on-site
and next-nearest-neighbor interactions,%

\begin{equation}
V_{c}(\vec{x}-\vec{x}^{\prime})=V_{c}^{(0)}\delta(\vec{x}-\vec{x}^{\prime
})+V_{c}^{(2)}\vec{\nabla}^{2}\delta(\vec{x}-\vec{x}^{\prime}),
\end{equation}

\begin{equation}
V_{\sigma}(\vec{x}-\vec{x}^{\prime})=V_{\sigma}^{(0)}\delta(\vec{x}-\vec
{x}^{\prime})+V_{\sigma}^{(2)}\vec{\nabla}^{2}\delta(\vec{x}-\vec{x}^{\prime
}).
\end{equation}
\qquad

The grand canonical partition function is given by%
\begin{equation}
Z=Tr\left[  \exp\left[  -\beta\left(  H-\mu_{\tau}n_{\tau}\right)  \right]
\right]  ,
\end{equation}
where $\mu_{\tau}$ is the isospin-dependent chemical potential and $n_{\tau}$
is the nucleon number operator for isospin index $\tau$. \ We can rewrite the
quartic interactions in $V_{c}$ and $V_{\sigma}$ using the
Hubbard-Stratonovich transformation \cite{Stratonovich:1958}%
\cite{Hubbard:1959ub}. \ The Hubbard-Stratonovich transformation uses the
identity,%
\begin{equation}
\exp(\tfrac{1}{2}A^{2})=\sqrt{2\pi}\int_{-\infty}^{\infty}d\varphi\exp
(-\tfrac{1}{2}\varphi^{2}-\varphi A),
\end{equation}
where $A$ is any quantum operator. \ This allows the mapping of the
interacting nucleon problem to a system of noninteracting nucleons coupled to
a fluctuating background field. \ With this transformation the expectation
value of any observable $O$ can be written as%
\begin{equation}
\left\langle O\right\rangle =\frac{\int D\phi G(\phi)\det(M(\phi))O(\phi
)}{\int D\phi G(\phi)\det(M(\phi))},
\end{equation}
where $\phi$ collectively represents the Hubbard-Stratonovich fields (as well
as any other bosonic fields), $M(\phi)$ is the one-body nucleon interaction
matrix, and $G(\phi)$ is a function of the $\phi$'s.

Using this formalism M\"{u}ller, \textit{et. al.}, were able to measure the
thermodynamic properties of nuclear matter and find signs of a first order
phase transition from an uncorrelated Fermi gas to a clustered phase. \ They
examined temperatures from 3.0 MeV to 100 MeV with up to 30 time steps and a
spatial volume of $4^{3}$ with lattice spacing $a=1.842$ fm$.$ \ Unfortunately
the LU factorization algorithm used to compute determinants in the simulation
scales as $(V\beta)^{3}$ and thus going beyond lattice systems of size $4^{3}$
is problematic.

\section{Chiral effective theory and nuclear forces}

There have been several recent efforts to describe nuclear forces starting
from chiral effective theory. \ This line of study was initiated by Weinberg
\cite{Weinberg:1990rz}\cite{Weinberg:1991um}\cite{Weinberg:1992yk}. \ The idea
is to expand the nuclear interactions in powers of $q/\Lambda_{\chi}$, where
$q$ is the typical external momentum of the nucleons and $\Lambda_{\chi}$ is
the chiral symmetry breaking scale or equivalently the hadronic mass scale
($\sim1$ GeV). \ The momentum cutoff scale $\Lambda$ for the effective theory
is set below $\Lambda_{\chi}$, and the renormalization group flow of operator
coefficients from scale $\Lambda_{\chi}$ to $\Lambda$ suppress the effects of
higher dimensional operators. \ Chiral symmetry in the limit of zero quark
mass imposes additional constraints on the possible momentum and spin
dependence of the interaction terms. \ Assuming naturalness for the
renormalized coupling constants in the Lagrangian at the scale $\Lambda_{\chi
}$, one expects in the low energy effective theory that contributions to
nucleon forces from operators at higher order in the chiral expansion are negligible.

Weinberg's work was followed by applications of chiral effective theory to the
nucleon potential \cite{Ordonez:1996rz} and alternative approaches to power
counting without apparent fine tuning in the presence of long scattering
lengths \cite{Kaplan:1998tg}\cite{Kaplan:1998we}. \ Recent low energy studies
\cite{Epelbaum:1998na}\cite{Epelbaum:1998hg}\cite{Epelbaum:1998ka}%
\cite{Epelbaum:2002vt} have also integrated out pion fields to produce energy
independent two- and three- nucleon potentials, and the effective theory
approach has been used to calculate nuclear spectra as well as phase shifts
and scattering lengths which compare favorably with potential model calculations.

In Weinberg's power counting scheme one deals with infrared singularities in
bound state problems by distinguishing between reducible and irreducible
diagrams. \ Reducible diagrams are those that can be disconnected by cutting
internal lines that correspond with particles in the initial or final state.
\ In the notation of \cite{Ordonez:1996rz}, the power of $q/\Lambda_{\chi}$
for any irreducible or non-reducible diagram is given by%
\begin{equation}
\nu=4-\tfrac{E_{f}}{2}+2L-2C+\sum_{i}V_{i}\delta_{i},
\end{equation}
where $E_{f}$ is the number of external nucleon lines, $L$ is the number of
loops, $C$ is the number of connected pieces, $V_{i}$ is the number of
vertices of type $i$, and $\delta_{i}$ is the index of vertex $i$. \ The index
$\delta_{i}$ is given by%
\begin{equation}
\delta_{i}=d_{i}+\tfrac{f_{i}}{2}-2\text{,}%
\end{equation}
where $d_{i}$ is the number of derivatives and $f_{i}$ is the number of
nucleon fields in the vertex.

We let $N$ represent the nucleon fields,%
\begin{equation}
N=\left[
\begin{array}
[c]{c}%
p\\
n
\end{array}
\right]  \otimes\left[
\begin{array}
[c]{c}%
\uparrow\\
\downarrow
\end{array}
\right]  .
\end{equation}
We use $\tau_{i}$ to represent Pauli matrices acting in isospin space, and we
use $\vec{\sigma}$ to represent Pauli matrices acting in spin space. Pion
fields are notated as $\pi_{i}$, and\ $\mu$ is the nucleon chemical potential.
\ We denote the pion decay constant as $F_{\pi}\approx183$ MeV and let
\begin{equation}
D=1+\pi_{i}^{2}/F_{\pi}^{2}.
\end{equation}
The lowest order Lagrange density for low energy pions and nucleons is given
by terms with $\delta_{i}=0$,
\begin{align}
\mathcal{L}^{(0)}  &  =-\tfrac{1}{2}D^{-2}\left[  (\vec{\nabla}\pi_{i}%
)^{2}-\dot{\pi}_{i}^{2}\right]  -\tfrac{1}{2}D^{-1}m_{\pi}^{2}\pi_{i}^{2}%
+\bar{N}[i\partial_{0}-(m_{N}-\mu)]N\nonumber\\
&  -D^{-1}F_{\pi}^{-1}g_{A}\bar{N}\left[  \tau_{i}\vec{\sigma}\cdot\vec
{\nabla}\pi_{i}\right]  N-D^{-1}F_{\pi}^{-2}\bar{N}[\epsilon_{ijk}\tau_{i}%
\pi_{j}\dot{\pi}_{k}]N\nonumber\\
&  -\tfrac{1}{2}C_{S}\bar{N}N\bar{N}N-\tfrac{1}{2}C_{T}\bar{N}\vec{\sigma
}N\cdot\bar{N}\vec{\sigma}N.
\end{align}
$g_{A}$ is the nucleon axial coupling constant, and $\epsilon_{ijk}$ is the
Levi-Civita symbol. For the purposes of power counting, the pion mass $m_{\pi
}$ is equivalent to one power of the momentum $q$. \ The nucleon mass term
$\bar{N}N$ actually has index $\delta_{i}=-1$. \ However the coefficient of
this term is fine-tuned using $\mu$ to set the nucleon density, and so the
$\bar{N}N$ term is reduced to the same size as other terms with index
$\delta_{i}=0$. \ At next order we have terms with $\delta_{i}=1$,%
\begin{equation}
\mathcal{L}^{(1)}=\tfrac{1}{2m_{N}}\bar{N}\vec{\nabla}^{2}N+...
\end{equation}
The important point for our discussion is that the lowest order Lagrange
density $\mathcal{L}^{(0)}$ describes static nucleons. \ Spatial hopping of
nucleons first appears at subleading index $\delta_{i}=1$. \ This suggests
that in some cases one could compute the determinant of the nucleon
interaction matrix as an expansion in powers of the spatial hopping parameter.
\ We should note one point of caution though. \ The $\mathcal{L}^{(1)}$
kinetic energy term cannot be ignored if the infrared singularities of
$\mathcal{L}^{(0)}$ are not properly dealt with. \ Since diagrammatic methods
are not used in nuclear matter Monte Carlo simulations, we cannot separate
reducible and irreducible diagrams. \ However if the simulation is done at
non-zero temperature $T$ then that will serve to regulate the infrared
singularities. \ We will explicitly see the effect of temperature on the
convergence of the expansion later in our discussion.

\section{Spatial zones}

In \cite{Lee:2003a} spatial zones were used to calculate the chiral condensate
for massless QED in three dimensions. \ In that paper the main problem was
dealing with sign and phase problems that arise in the Hamiltonian worldline
formalism with explicit fermions. \ The idea of the zone method is that
fermions at inverse temperature $\beta$ with spatial hopping parameter $h$
have a localization length of%
\begin{equation}
l\sim\sqrt{\beta h}.
\end{equation}
We now apply the zone idea to our determinant problem. \ Let $M$ be the
nucleon matrix, in general an $n\times n$ complex matrix. \ We partition the
lattice spatially into separate zones such that the length of each zone is
much larger than the localization length $l$. \ Since most nucleon worldlines
do not cross the zone boundaries, they would not be affected if we set the
zone boundary hopping terms to zero. \ Hence we anticipate that the
determinant of $M$ can be approximated by the product of the submatrix
determinants for each spatial zone.

Let us partition the lattice into spatial zones labelled by index $j.$ \ Let
$\{P_{j}\}$ be a complete set of matrix projection operators that project onto
the lattice sites within spatial zone $j$. \ We can write%
\begin{equation}
M=\sum_{i,j}P_{j}MP_{i}=M_{0}+M_{E},
\end{equation}
where%
\begin{align}
M_{0}  &  =\sum_{i}P_{i}MP_{i},\\
M_{E}  &  =\sum_{i\neq j}P_{j}MP_{i}.
\end{align}
If the zones can be sorted into even and odd sets so that%
\begin{equation}
P_{j}MP_{i}=0
\end{equation}
whenever $i$ is even and $j$ is odd or vice-versa, then we say that the zone
partitioning is bipartite. \ We now have%
\begin{align}
\det(M)  &  =\det(M_{0})\det(1+M_{0}^{-1}M_{E})\nonumber\\
&  =\det(M_{0})\exp(\text{trace}(\log(1+M_{0}^{-1}M_{E})))\text{.}%
\end{align}
Using an expansion for the logarithm, we have%
\begin{equation}
\det(M)=\det(M_{0})\exp\left(  \sum_{p=1}^{\infty}\frac{(-1)^{p-1}}%
{p}\text{trace}((M_{0}^{-1}M_{E})^{p})\right)  .
\end{equation}

Let us define
\begin{equation}
\Delta_{m}=\det(M_{0})\exp\left(  \sum_{p=1}^{m}\frac{(-1)^{p-1}}%
{p}\text{trace}((M_{0}^{-1}M_{E})^{p})\right)  .
\end{equation}
Let $\lambda_{k}(M_{0}^{-1}M_{E})$ be the eigenvalues of $M_{0}^{-1}M_{E}$ and
$R$ be the spectral radius,%
\begin{equation}
R=\max_{k=1,...,n}(\left\vert \lambda_{k}(M_{0}^{-1}M_{E})\right\vert ).
\end{equation}
It has been shown \cite{Ipsen:2003} that for $R<1$%
\begin{equation}
\frac{\left\vert \det(M)-\Delta_{m}\right\vert }{\left\vert \Delta
_{m}\right\vert }\leq cR^{m}e^{cR^{m}} \label{bound}%
\end{equation}
where%
\begin{equation}
c=-n\log(1-R). \label{prefactor}%
\end{equation}
The spectral radius $R$ determines the convergence of our expansion. \ $R$ can
be reduced by increasing the size of the spatial zone relative to the
localization length $l$. \ In the special case where the zone partitioning is
bipartite, we note that for any odd $p,$%
\begin{equation}
\text{trace}((M_{0}^{-1}M_{E})^{p})=0.
\end{equation}
In that case
\begin{equation}
\Delta_{2m+1}=\Delta_{2m},
\end{equation}
and so we gain an extra order of accuracy without additional work.

\section{Application to neutron matter simulations}

We now illustrate the zone determinant expansion for a simple but realistic
lattice simulation of neutron matter. \ The formalism we use is a merger of
chiral effective theory and Euclidean lattice methods. \ In our analysis we
focus on the convergence and accuracy of the zone determinant expansion
method. \ In order to provide a detailed quantitative assessment we compare
the zone determinant approximations with exact determinant results. \ Given
the severe limitations of the exact determinant method, we are not able to
probe large volumes nor comment on finite volume scaling. \ However a full
account of results for large volume simulations as well as an introduction to
the new approach combining chiral effective theory and lattice methods is
forthcoming in another paper \cite{Borasoy:2003}.

Our starting point is the same as that of Weinberg \cite{Weinberg:1990rz}.
\ We start with the most general local Lagrange density involving pions and
low-energy nucleons consistent with Lorentz and translational invariance,
isospin symmetry, and spontaneously broken chiral symmetry. \ This will
produce an infinite set of interaction terms with increasing numbers of
derivatives and/or nucleon fields. \ The dependence of each term on the pion
field is governed by the rules of spontaneously broken chiral symmetry.
\ Degrees of freedom associated with heavier mesons such as the $\rho$,
heavier baryons such as the $\Delta$'s as well as antinucleons are all
integrated out. \ We also integrate out nucleons with momenta greater than the
cutoff scale $\Lambda$. \ The contribution of these effects appear as
coefficients of local terms in our pion-nucleon Lagrangian.

For simplicity we consider only neutrons and neutral pions. \ We let $\psi$
represent the neutron spin states,%
\begin{equation}
\psi=\left[
\begin{array}
[c]{c}%
\uparrow\\
\downarrow
\end{array}
\right]  .
\end{equation}
The terms in our effective pion-nucleon Euclidean action are%

\begin{equation}
S=S_{\pi\pi}+S_{\bar{N}N}+S_{\pi\bar{N}N}+S_{\bar{N}N\bar{N}N},
\end{equation}
where%
\begin{equation}
S_{\pi\pi}=%
{\textstyle\int}
d^{3}\vec{r}dr_{4}\left[  \tfrac{1}{2}(\tfrac{\partial\pi_{0}}{\partial r_{4}%
})^{2}+\tfrac{1}{2}(\vec{\nabla}\pi_{0})^{2}+\tfrac{1}{2}m_{\pi}^{2}\pi
_{0}^{2}\right]  ,
\end{equation}%
\begin{equation}
S_{\bar{N}N}=%
{\textstyle\int}
d^{3}\vec{r}dr_{4}\left[  \psi^{\dagger}\tfrac{\partial\psi}{\partial r_{4}%
}-\psi^{\dagger}\tfrac{\vec{\nabla}^{2}\psi}{2m_{N}}+(m_{N}-\mu)\psi^{\dagger
}\psi\right]  ,
\end{equation}%
\begin{equation}
S_{\pi\bar{N}N}=%
{\textstyle\int}
d^{3}\vec{r}dr_{4}\left[  -g\psi^{\dagger}\left(  \vec{\sigma}\cdot\vec
{\nabla}\pi_{0}\right)  \psi\right]  ,
\end{equation}%
\begin{equation}
S_{\bar{N}N\bar{N}N}=%
{\textstyle\int}
d^{3}\vec{r}dr_{4}\left[  \tfrac{1}{2}C\text{:}\psi^{\dagger}\psi\psi
^{\dagger}\psi\text{:}\right]  . \label{fourfermi}%
\end{equation}
We have kept terms in the lowest order chiral Lagrange density $\mathcal{L}%
^{(0)}$ containing neutrons and neutral pions. \ We have dropped the factors
of $D^{-1}$ which at lowest order contribute to multipion processes. \ We have
also chosen to include the neutron kinetic energy term even though it appears
in $\mathcal{L}^{(1)}$. \ This is useful if we wish to recover the exact free
neutron Fermi gas in the weak coupling limit.

On the lattice we let $a$ be the spatial lattice spacing and $a_{t}$ be the
temporal lattice spacing. \ We use the notation $\hat{1},\hat{2},\hat{3},$ or
$\hat{4}$ to represent vectors that extend exactly one lattice unit (either
$a$ or $a_{t}$) in the respective direction. \ We use dimensionless lattice
fields and dimensionless masses and couplings by multiplying by the
corresponding power of $a$. \ For example, $\hat{m}_{\pi}=m_{\pi}\cdot a$,
$\hat{m}_{N}=m_{N}\cdot a$, $\hat{g}=g\cdot a^{-1}$, $\hat{\mu}=\mu\cdot a$,
and $\hat{C}=C\cdot a^{-2}$. \ We use standard fermion path integral
conventions at finite time step \cite{Soper:1978dp}\cite{Rothe:1997kp} and
define%
\begin{equation}
\psi^{\prime}(\vec{n})=\psi(\vec{n}-\hat{4})
\end{equation}
in order to write the lattice path integral in standard form. \ The lattice
actions have the form%
\begin{align}
S_{\pi\pi}  &  =-\tfrac{a}{a_{t}}\sum_{\vec{n}}\pi(\vec{n})\pi(\vec{n}+\hat
{4})-\tfrac{a_{t}}{a}\sum_{\vec{n},l=1,2,3}\pi(\vec{n})\pi(\vec{n}+\hat
{l})\nonumber\\
&  +((\tfrac{\hat{m}_{\pi}^{2}}{2}+3)\tfrac{a_{t}}{a}+\tfrac{a}{a_{t}}%
)\sum_{\vec{n}}(\pi(\vec{n}))^{2},
\end{align}%
\begin{align}
S_{\bar{N}N}  &  =\sum_{\vec{n}}\psi^{\dagger}(\vec{n})\psi^{\prime}(\vec
{n}+\hat{4})-\tfrac{a_{t}}{2\hat{m}_{N}a}\sum_{\vec{n},l=1,2,3}(\psi^{\dagger
}(\vec{n})\psi^{\prime}(\vec{n}+\hat{l})+\psi^{\dagger}(\vec{n})\psi^{\prime
}(\vec{n}-\hat{l}))\nonumber\\
&  +(-1+(\hat{m}_{N}+\tfrac{3}{\hat{m}_{N}})\tfrac{a_{t}}{a}-\hat{\mu}%
\tfrac{a_{t}}{a})\sum_{\vec{n}}\psi^{\dagger}(\vec{n})\psi^{\prime}(\vec{n}),
\end{align}%
\begin{align}
S_{\pi\bar{N}N}  &  =-\tfrac{\hat{g}a_{t}}{2a}%
{\displaystyle\sum_{\vec{n}}}
\left[  \left[  \psi_{\uparrow}^{\ast}(\vec{n})\psi_{\uparrow}^{\prime}%
(\vec{n})-\psi_{\downarrow}^{\ast}(\vec{n})\psi_{\downarrow}^{\prime}(\vec
{n})\right]  \left[  \pi(\vec{n}+\hat{3})-\pi(\vec{n}-\hat{3})\right]  \right]
\nonumber\\
&  -\tfrac{\hat{g}a_{t}}{2a}%
{\displaystyle\sum_{\vec{n}}}
\left[  \psi_{\uparrow}^{\ast}(\vec{n})\psi_{\downarrow}^{\prime}(\vec
{n})\left[  \pi(\vec{n}+\hat{1})-\pi(\vec{n}-\hat{1})-i\pi(\vec{n}+\hat
{2})+i\pi(\vec{n}-\hat{2})\right]  \right] \nonumber\\
&  -\tfrac{\hat{g}a_{t}}{2a}%
{\displaystyle\sum_{\vec{n}}}
\left[  \psi_{\downarrow}^{\ast}(\vec{n})\psi_{\uparrow}^{\prime}(\vec
{n})\left[  \pi(\vec{n}+\hat{1})-\pi(\vec{n}-\hat{1})+i\pi(\vec{n}+\hat
{2})-i\pi(\vec{n}-\hat{2})\right]  \right]  ,
\end{align}%
\begin{equation}
S_{\bar{N}N\bar{N}N}=\tfrac{\hat{C}a_{t}}{a}\sum_{\vec{n}}\psi_{\uparrow
}^{\ast}(\vec{n})\psi_{\uparrow}^{\prime}(\vec{n})\psi_{\downarrow}^{\ast
}(\vec{n})\psi_{\downarrow}^{\prime}(\vec{n}).
\end{equation}
We can reexpress $S_{\bar{N}N\bar{N}N}$ using a discrete Hubbard-Stratonovich
transformation \cite{Hirsch:1983} for $\hat{C}\leq0$,
\begin{align}
&  \exp(-\tfrac{\hat{C}a_{t}}{a}\psi_{\uparrow}^{\ast}(\vec{n})\psi_{\uparrow
}^{\prime}(\vec{n})\psi_{\downarrow}^{\ast}(\vec{n})\psi_{\downarrow}^{\prime
}(\vec{n}))\nonumber\\
&  =\tfrac{1}{2}\sum_{s=\pm1}\exp\left[  -(\tfrac{\hat{C}a_{t}}{2a}+\lambda
s)(\psi_{\uparrow}^{\ast}(\vec{n})\psi_{\uparrow}^{\prime}(\vec{n}%
)+\psi_{\downarrow}^{\ast}(\vec{n})\psi_{\downarrow}^{\prime}(\vec
{n})-1)\right]  ,
\end{align}
where $\lambda$ is defined by%
\begin{equation}
\cosh\lambda=\exp(-\tfrac{\hat{C}a_{t}}{2a}).
\end{equation}

\section{Results}

For the simulation results presented in this section we use the values
$a^{-1}=150$\ MeV, $a_{t}^{-1}=225$\ MeV, $C=-4.0\cdot10^{-5}\ $MeV$^{-2}$,
and $g=6.8\cdot10^{-3}$ MeV$^{-1}$. \ The value of $g$ is set according to the
tree level Goldberger-Treiman relation \cite{Goldberger:1958vp}%
\begin{equation}
g=F_{\pi}^{-1}g_{A}=6.8\cdot10^{-3}\text{ MeV}^{-1}.
\end{equation}
The value of $C$ is tuned to match the $s$-wave phase shifts on the lattice
for nucleon scattering at lattice spacing $(150$\ MeV$)^{-1}$. \ The
calculation of phase shifts on the lattice is discussed in a forthcoming
article which details the entire formalism \cite{Borasoy:2003}.

We present data for three independent pion and Hubbard-Stratonovich field
configurations on a $4^{3}\times6$ lattice at temperature $T=37.5$ MeV and
neutron density $\rho=$ 0.57$\rho_{\text{nucl}}$ where nuclear density is
\begin{equation}
\rho_{\text{nucl}}=2.8\cdot10^{14}\text{ g/cm}^{3}=1.2\cdot10^{9}\text{
MeV}^{4}\text{.}%
\end{equation}
We refer to spatial zones according to their $x,y,z$ lattice dimensions
$[n_{x},n_{y},n_{z}].$ \ At this lattice spacing and temperature our
localization length estimate is%
\begin{equation}
l\sim\sqrt{\beta h}=0.57,
\end{equation}
and so we expect the zone determinant expansion to converge for even the
smallest zones, $[n_{x},n_{y},n_{z}]=[1,1,1],$ consisting of groups of lattice
points with the same spatial coordinate. \ In Table 1 we show the spectral
radius $R$ of $M_{0}^{-1}M_{E}$ and the determinant expansion for $[1,1,1]$
zones for three independent pion and Hubbard-Stratonovich configurations at equilibrium.%

\[%
\begin{tabular}
[c]{||l|l|l|l||}\hline\hline
configuration & $\#1$ & $\#2$ & $\#3$\\\hline
$R$ & $0.538$ & $0.523$ & $0.534$\\\hline
$\log(\det(M))$ & $16.4727+0.3666i$ & $18.1193+0.4479i$ & $18.2612-0.1811i$%
\\\hline
$\log(\Delta_{0})$ & $12.1700+0.3807i$ & $13.7933+0.4900i$ & $14.0060-0.2075i$%
\\\hline
$\log(\Delta_{2})$ & $16.4964+0.3686i$ & $18.1792+0.4477i$ & $18.2841-0.1801i$%
\\\hline
$\log(\Delta_{4})$ & $16.4959+0.3659i$ & $18.1329+0.4468i$ & $18.2856-0.1805i$%
\\\hline
$\log(\Delta_{6})$ & $16.4664+0.3667i$ & $18.1147+0.4482i$ & $18.2548-0.1813i$%
\\\hline
$\log(\Delta_{8})$ & $16.4739+0.3666i$ & $18.1203+0.4478i$ & $18.2622-0.1810i$%
\\\hline\hline
\end{tabular}
\]

\begin{center}
Table 1: \ Convergence of determinant series for $[1,1,1]$ zones
\end{center}

We see that the spectral radius $R$ is less than $1$ as expected from the
localization length estimate. \ We also find that the rigorous bounds in
(\ref{bound}) are satisfied. \ Since most of the eigenvalues of $M_{0}%
^{-1}M_{E}$ are much smaller in magnitude than the spectral radius $R$, we
observe that the prefactor $c$ in (\ref{bound}) is actually much larger than
needed for these examples. \ From the data in Table 1 we find empirically%
\begin{equation}
\frac{\left\vert \det(M)-\Delta_{2m}\right\vert }{\left\vert \Delta
_{2m}\right\vert }=\frac{\left\vert \det(M)-\Delta_{2m+1}\right\vert
}{\left\vert \Delta_{2m+1}\right\vert }\sim R^{2m+1}.
\end{equation}

We now repeat the same analysis with the same lattice dimensions and
temperature but at much higher density, $\rho=$ 1.67$\rho_{\text{nucl}}$.
\ The results for three independent pion and Hubbard-Stratonovich field
configurations are shown in Table 2.%
\[%
\genfrac{}{}{0pt}{}{%
\begin{tabular}
[c]{||l|l|l|l||}\hline\hline
configuration & $\#1$ & $\#2$ & $\#3$\\\hline
$R$ & $0.520$ & $0.478$ & $0.536$\\\hline
$\log(\det(M))$ & $70.6349+0.4121i$ & $76.1762+0.7098i$ & $73.4831-0.1576i$%
\\\hline
$\log(\Delta_{0})$ & $64.8000+0.4406i$ & $71.0009+0.7989i$ & $67.8503-0.1809i$%
\\\hline
$\log(\Delta_{2})$ & $70.9530+0.4120i$ & $76.3927+0.7005i$ & $73.7923-0.1538i$%
\\\hline
$\log(\Delta_{4})$ & $70.6032+0.4114i$ & $76.1605+0.7106i$ & $73.4483-0.1583i$%
\\\hline
$\log(\Delta_{6})$ & $70.6390+0.4124i$ & $76.1778+0.7097i$ & $73.4888-0.1574i$%
\\\hline
$\log(\Delta_{8})$ & $70.6343+0.4120i$ & $76.1760+0.7098i$ & $73.4819-0.1576i$%
\\\hline\hline
\end{tabular}
}{\text{Table 2:\ \ Determinant series for [1,1,1] zones at }\rho
=1.67\rho_{\text{nucl}}}%
\]
Comparing Tables 1 and 2, we conclude that the zone determinant expansion
appears to be unaffected by the increase in nucleon density.

On the other hand as the temperature decreases, the localization length
increases and therefore the convergence of the zone determinant expansion
should slow down. \ In Table 3 we show the expansion for a $6^{3}\times6$
lattice at $T=37.5$ MeV, $6^{3}\times9$ lattice at $T=25.0$ MeV, $6^{3}%
\times12$ lattice at $T=18.8$ MeV, $6^{3}\times18$ lattice at $T=12.5$ MeV.
\ The chemical potential is kept at the same value used for the results in
Table 1 ($\mu=0.8m_{N})$.%

\[%
\begin{tabular}
[c]{||l|l|l|l|l||}\hline\hline
$T$ & $37.5$ MeV & $25.0$ MeV & $18.8$ MeV & $12.5$ MeV\\\hline
$R$ & $0.5122$ & $0.6742$ & $0.7631$ & $0.8638$\\\hline
$\log(\det(M))$ & $65.8009-0.7250i$ & $37.6912-1.5930i$ & $16.0023+0.0541i$ &
$5.7618+0.0452i$\\\hline
$\log(\Delta_{0})$ & $51.3988-0.7888i$ & $20.7653-1.5752i$ & $4.1999+0.2033i$
& $0.1796-0.0019i$\\\hline
$\log(\Delta_{2})$ & $66.0028-0.7218i$ & $36.2570-1.6535i$ & $11.7590+0.0861i$
& $1.5071+0.0222i$\\\hline
$\log(\Delta_{4})$ & $65.8289-0.7243i$ & $38.0872-1.5800i$ & $15.9390+0.0337i$
& $3.8579+0.0399i$\\\hline
$\log(\Delta_{6})$ & $65.7926-0.7253i$ & $37.6528-1.5924i$ & $16.3121+0.0478i$
& $5.5534+0.0475i$\\\hline
$\log(\Delta_{8})$ & $65.8026-0.7249i$ & $37.6780-1.5943i$ & $16.0111+0.0575i$
& $6.0067+0.0473i$\\\hline\hline
\end{tabular}
\
\]

\begin{center}
Table 3: \ Determinant series for $[1,1,1]$ zones for varying temperatures
\end{center}

\noindent The increase in $R$ and the slowdown in the series convergence is
consistent with our intuition based on the localization length. \ Although we
are looking at neutron matter in this example, we comment that this is the
appropriate temperature range for observing the liquid-gas transition in
symmetric nuclear matter ($T\sim20$ MeV) \cite{Elliott:2001hn}\cite{Ma:2003dc}%
\cite{Karnaukhov:2003vp}. \ The localization length becomes greater than $1$
for $T\sim10$ MeV, and so we expect the series to break down for $[1,1,1]$
zones at colder temperatures. \ This also happens to be the temperature at
which long range interactions due to pairing become important. \ In order to
see pairing correlations, one should therefore use larger zone sizes.
\ However we note that at temperatures much less than $10$ MeV there is also a
significant sign problem, and the numerical difficulties will be substantial
independent of the method used to calculate determinants.

We now look at how convergence of the expansion is improved by using larger
spatial zones. \ In Table 4 we show the determinant expansion for a
$6^{3}\times6$ lattice at $T=37.5$ MeV for $[1,1,1]$, $[2,2,2],$ and $[3,3,3]$
zones \footnote{For $[2,2,2]$ the zone breakup is no longer biparitite since
the $6\div2$ is odd. \ However this is only a boundary effect and we find that
$\Delta_{2m+1}$ is extremely close to $\Delta_{2m}$.}.%
\[%
\begin{tabular}
[c]{||l|l|l|l||}\hline\hline
zone & $[1,1,1]$ & $[2,2,2]$ & $[3,3,3]$\\\hline
$R$ & $0.5122$ & $0.3013$ & $0.3014$\\\hline
$\log(\det(M))$ & $65.8009-0.7250i$ & $65.8009-0.7250i$ & $65.8009-0.7250i$%
\\\hline
$\log(\Delta_{0})$ & $51.3988-0.7888i$ & $58.6585-0.7543i$ & $61.0471-0.7722i$%
\\\hline
$\log(\Delta_{2})$ & $66.0028-0.7218i$ & $65.8570-0.7231i$ & $65.8317-0.7233i$%
\\\hline
$\log(\Delta_{4})$ & $65.8289-0.7243i$ & $65.8015-0.7251i$ & $65.8007-0.7250i$%
\\\hline
$\log(\Delta_{6})$ & $65.7926-0.7253i$ & $65.8009-0.7250i$ & $65.8009-0.7250i$%
\\\hline
$\log(\Delta_{8})$ & $65.8026-0.7249i$ & $65.8009-0.7250i$ & $65.8009-0.7250i$%
\\\hline\hline
\end{tabular}
\ \
\]

\begin{center}
Table 4: \ Determinant series for various zone sizes
\end{center}

\noindent It is somewhat unusual that the spectral radius $R$ is about the
same for $[2,2,2]$ and $[3,3,3]$, however the convergence of the series
clearly improves as we increase the zone size.

We now investigate the convergence of the determinant zone expansion for
physical observables. \ Let us return to the data in Table 1 for a moment.
\ We observe that at any given order the error appears to be about the same
for each of the three independent configurations. \ Since the measurement of a
physical observable does not dependent on the overall normalization of the
partition function, this suggests that the zone determinant expansion could be
more accurate in approximating physical observables. \ We now check to see if
this is so for a particular example.

Let us define the neutron occupation number at site $\vec{r}$,
\begin{equation}
\rho_{\uparrow(\downarrow)}(\vec{r})=\psi_{\uparrow(\downarrow)}^{\ast}%
(\vec{r})\psi_{\uparrow(\downarrow)}(\vec{r})=\psi_{\uparrow(\downarrow
)}^{\ast}(\vec{r})\psi_{\uparrow(\downarrow)}^{\prime}(\vec{r}+\hat{4}).
\end{equation}
We measure the opposite spin radial distribution function,%
\begin{equation}
<\rho_{\uparrow}(\vec{r})\rho_{\downarrow}(0)>,
\end{equation}
by sampling $20$ independent pion and Hubbard-Stratonovich field
configurations. \ For $r_{x}=0$ we plot the results for the opposite spin
radial distribution function in Fig. 1 using exact matrix determinants for a
$4^{3}\times6$ lattice at $T=37.5$ MeV. \ In Figs. 2-6 we show the error in
the radial distribution function if the estimate $\Delta_{m}$ is used in place
of $\det(M)$ for $m=0,2,4,6,8$.%

\begin{figure}
[ptb]
\begin{center}
\includegraphics[
height=3.3183in,
width=4.062in
]%
{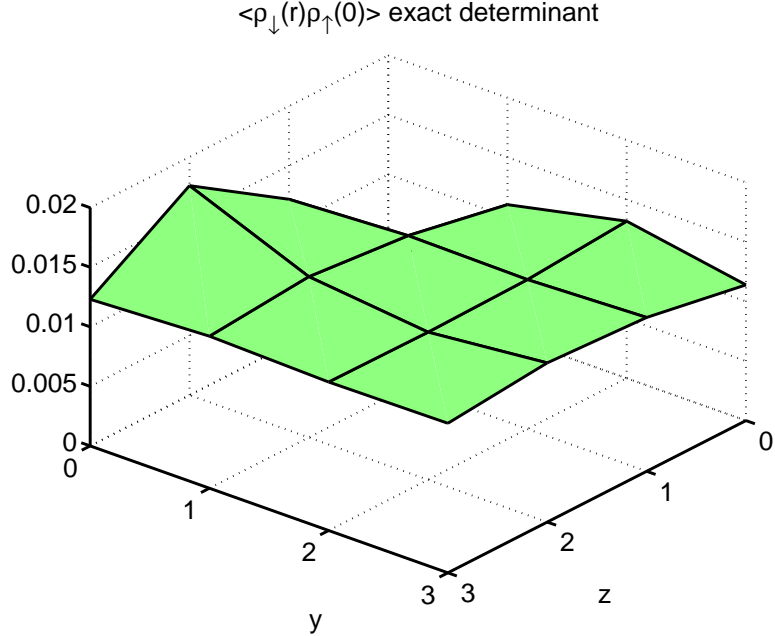}%
\caption{Opposite spin radial distribution function using exact determinants.}%
\label{exact}%
\end{center}
\end{figure}
\begin{figure}
[ptbptb]
\begin{center}
\includegraphics[
height=3.3183in,
width=3.8778in
]%
{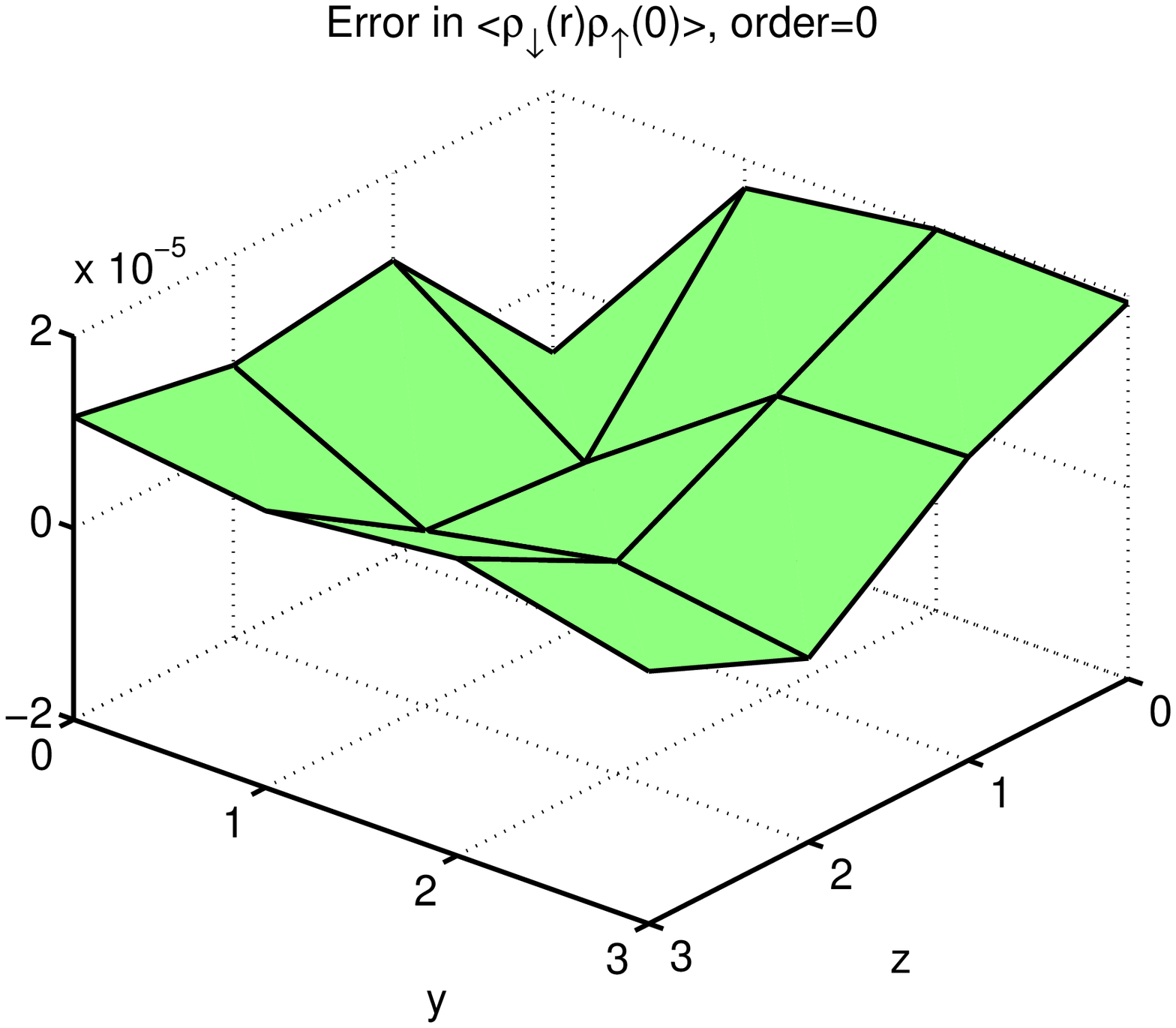}%
\caption{Error in the opposite spin radial distribution function at order
$m=0$.}%
\label{order0}%
\end{center}
\end{figure}
\begin{figure}
[ptbptbptb]
\begin{center}
\includegraphics[
height=3.3183in,
width=3.9081in
]%
{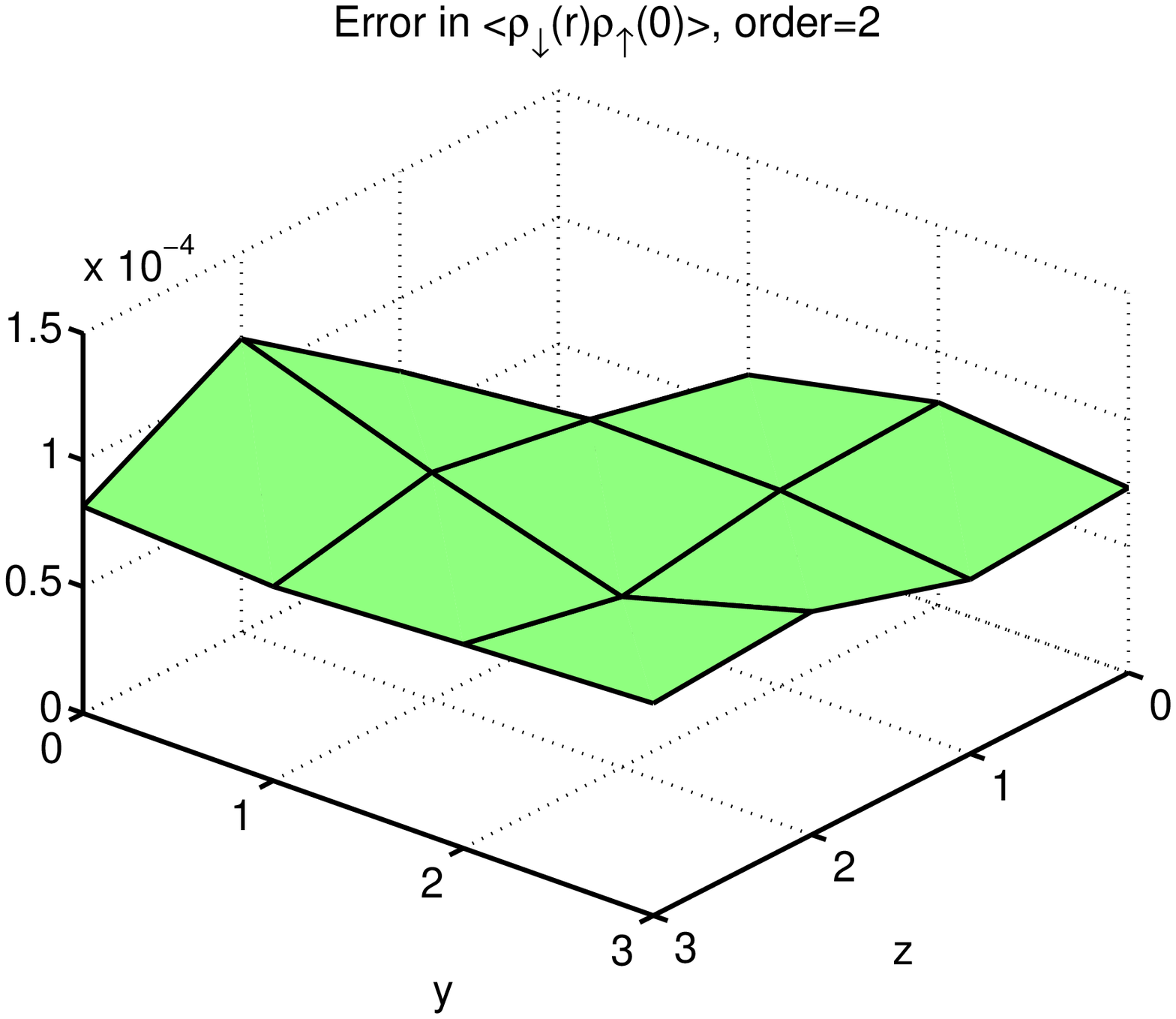}%
\caption{Error in the opposite spin radial distribution function at order
$m=2$.}%
\label{order2}%
\end{center}
\end{figure}
\begin{figure}
[ptbptbptbptb]
\begin{center}
\includegraphics[
height=3.3183in,
width=3.8778in
]%
{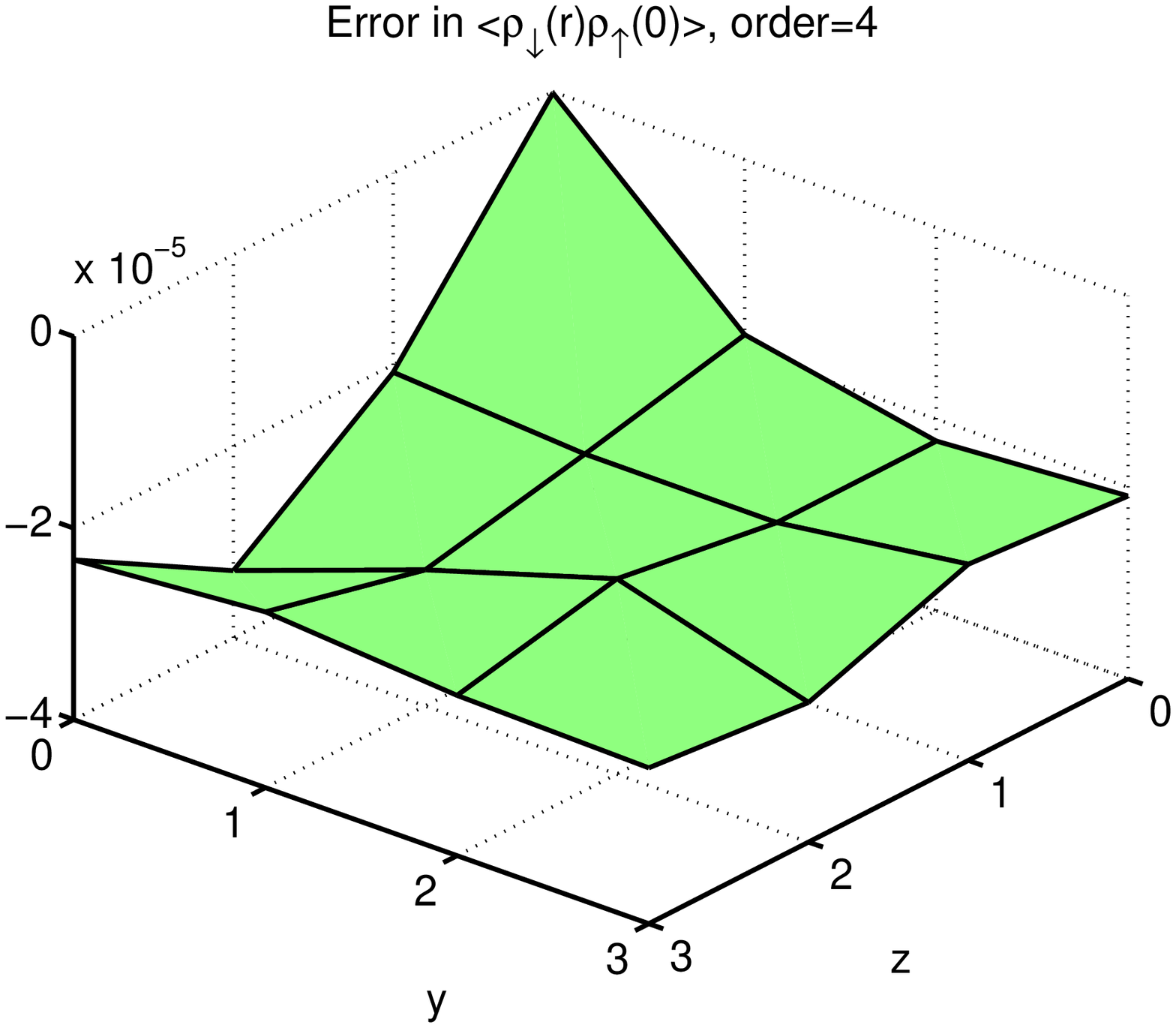}%
\caption{Error in the opposite spin radial distribution function at order
$m=4$.}%
\label{order4}%
\end{center}
\end{figure}
\begin{figure}
[ptbptbptbptbptb]
\begin{center}
\includegraphics[
height=3.3183in,
width=3.9081in
]%
{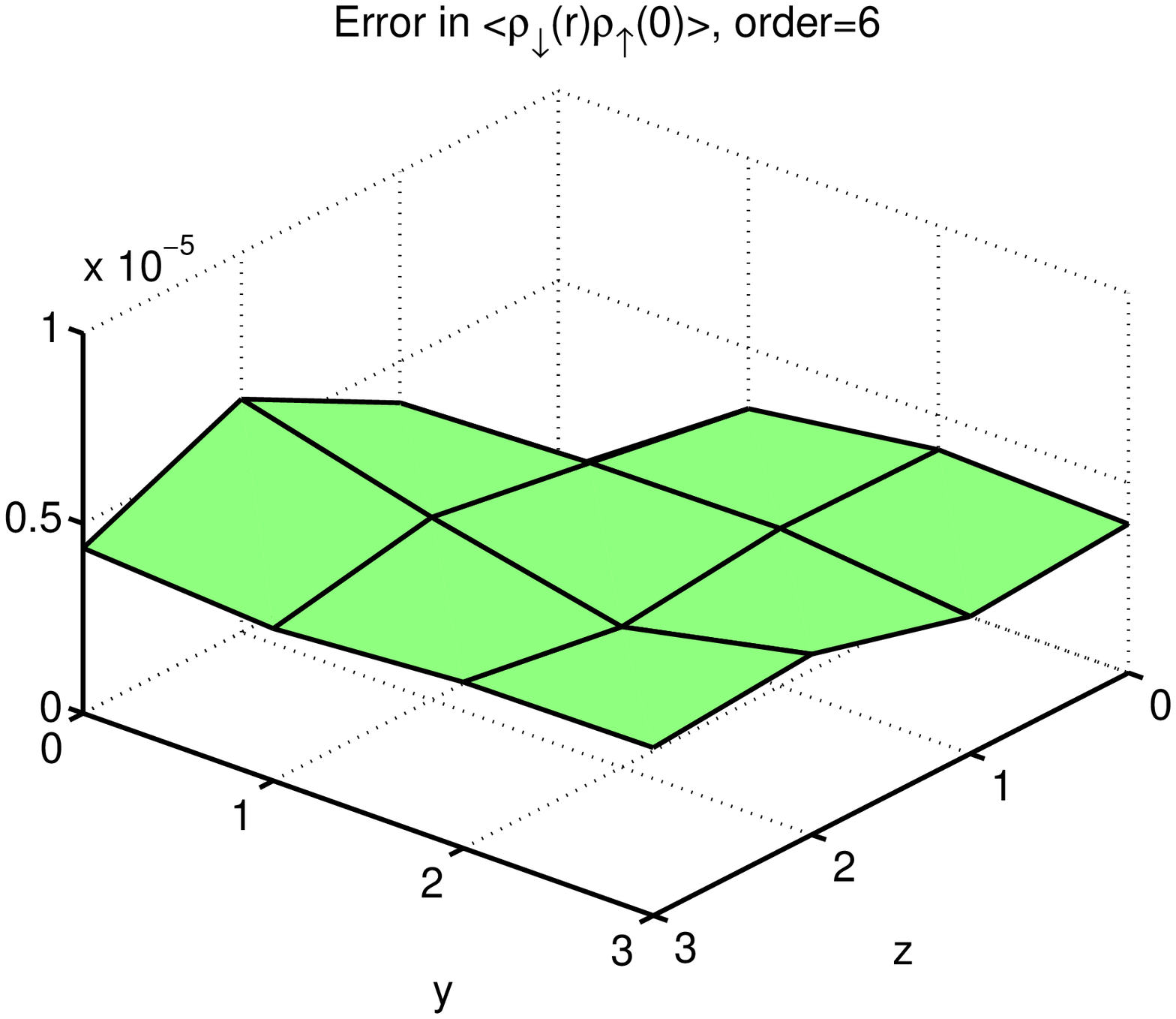}%
\caption{Error in the opposite spin radial distribution function at order
$m=6$.}%
\label{order6}%
\end{center}
\end{figure}
\begin{figure}
[ptbptbptbptbptbptb]
\begin{center}
\includegraphics[
height=3.3183in,
width=3.8778in
]%
{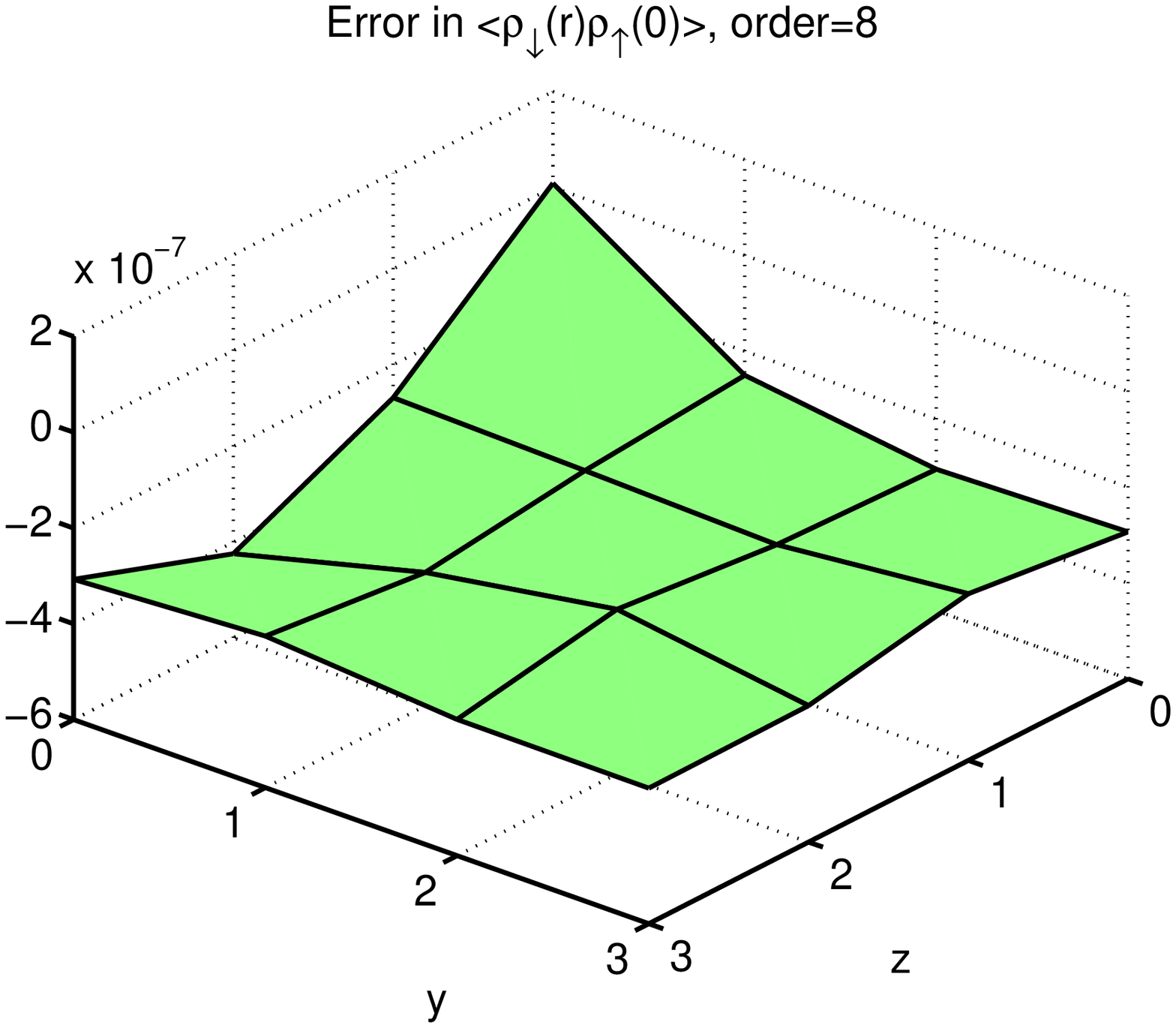}%
\caption{Error in the opposite spin radial distribution function at order
$m=8$.}%
\label{order8}%
\end{center}
\end{figure}

The approximation at order $0$ is a lot better than expected given the error
of $\Delta_{0}$ in approximating $\det(M)$. \ Overall we find that the zone
expansion is significantly more accurate for the radial distribution function
than the expansion of the determinants. \ It is premature to say if this is
typical of all physical observable measurements. \ Nevertheless if most of the
error of $\Delta_{m}$ is in fact independent of the pion and
Hubbard-Stratonovich configurations at equilibrium, then we expect an
improvement in accuracy for most physical observables.

\section{Summary and conclusions}

We have discussed lattice simulations of finite temperature nuclear matter and
a new approximation method called the zone determinant expansion for nucleon
matrix determinants. \ The expansion is made possible by the small size of the
spatial hopping parameter. \ We know from power counting in chiral effective
theory that the spatial hopping parameter is suppressed relative to the
leading order interactions at low energies. \ The zone determinant expansion
is given by%
\begin{equation}
\det(M)=\det(M_{0})\exp\left(  \sum_{p=1}^{\infty}\frac{(-1)^{p-1}}%
{p}\text{trace}((M_{0}^{-1}M_{E})^{p})\right)  ,
\end{equation}
where $M$ is the nucleon one-body interaction matrix, $M_{E}$ is the submatrix
consisting of zone boundary hopping terms, and $M_{0}$ is the submatrix
without boundary hopping terms. \ The convergence of the expansion is
controlled by the spectral radius of $M_{0}^{-1}M_{E}$. \ Physically we expect
the convergence to be rapid if the localization length of the nucleons%
\begin{equation}
l\sim\sqrt{\beta h}%
\end{equation}
is small compared to the size of the spatial zones.

We tested the zone determinant expansion using lattice simulations of neutron
matter with self interactions and neutral pion exchange. \ The convergence of
the expansion was measured for several configurations at temperature $T=37.5$
MeV and using $[1,1,1]$ spatial zones. \ By decreasing the temperature from
$T=37.5$ MeV to $12.5$ MeV we found that the convergence of the expansion
becomes slower, as predicted by the increase in the localization length. \ But
we then showed that convergence could be accelerated by increasing the size of
the zones from $[1,1,1]$ to $[3,3,3]$. \ Finally we looked at the convergence
of the expansion for the opposite spin radial distribution function%
\begin{equation}
<\rho_{\uparrow}(\vec{r})\rho_{\downarrow}(0)>.
\end{equation}
We found that the accuracy of the expansion for this physical observable was
significantly better at each order than that for the expansion of the determinants.

The number of required operations for calculating the nucleon determinant
using LU factorization for an $n\times n$ matrix scales as $n^{3}$.
\ Therefore a nuclear lattice simulation that includes nucleon/nucleon-hole
loops requires $(V\beta)^{3}$ times more operations than the quenched
simulation without loops. \ This numerical challenge has been the most
pressing limitation on finite temperature nuclear lattice simulations to date.

For the zone determinant expansion method at fixed zone size, the computation
cost scales only as $f(m)\beta^{3}$ where $m$ is the order of the expansion.
\ For a simulation on a lattice with spatial dimensions $8^{3}$, one can
accelerate the simulation by a factor of about $10^{5}$ to $10^{7}$, depending
on the expansion order and size of the spatial zone. \ The savings are greater
on larger lattices and should facilitate future work in the area of finite
temperature nuclear lattice simulations.

\begin{acknowledgments}
D.L. is grateful to B. Borasoy, T. Schaefer, R. Seki, U. van Kolck, and
participants at the 2003 CECAM Sign Problem Workshop for discussions and
comments. \ This work was supported in part by NSF Grants DMS-0209931 and DMS-0209695.
\end{acknowledgments}

\bibliographystyle{h-physrev3}
\bibliography{NuclearMatter}

\begin{thebibliography}{10}

\bibitem{Scalettar:1986uy}
R.~T. Scalettar, D.~J. Scalapino, and R.~L. Sugar,
\newblock Phys. Rev. {\bf B34}, 7911 (1986).

\bibitem{Gottlieb:1987mq}
S.~Gottlieb, W.~Liu, D.~Toussaint, R.~L. Renken, and R.~L. Sugar,
\newblock Phys. Rev. {\bf D35}, 2531 (1987).

\bibitem{Duane:1987de}
S.~Duane, A.~D. Kennedy, B.~J. Pendleton, and D.~Roweth,
\newblock Phys. Lett. {\bf B195}, 216 (1987).

\bibitem{Scalapino:1981qs}
D.~J. Scalapino and R.~L. Sugar,
\newblock Phys. Rev. Lett. {\bf 46}, 519 (1981).

\bibitem{Gubernatis:1992a}
E.~Y.~J. Loh and J.~E. Gubernatis,
\newblock {\em Electronic Phase Transitions} (Elsevier Science Publishers,
  1992), chap.~4, pp. 177--235.

\bibitem{Muller:1999cp}
H.~M. Muller, S.~E. Koonin, R.~Seki, and U.~van Kolck,
\newblock Phys. Rev. {\bf C61}, 044320 (2000), nucl-th/9910038.

\bibitem{Stratonovich:1958}
R.~L. Stratonovich,
\newblock Soviet Phys. Doklady {\bf 2}, 416 (1958).

\bibitem{Hubbard:1959ub}
J.~Hubbard,
\newblock Phys. Rev. Lett. {\bf 3}, 77 (1959).

\bibitem{Weinberg:1990rz}
S.~Weinberg,
\newblock Phys. Lett. {\bf B251}, 288 (1990).

\bibitem{Weinberg:1991um}
S.~Weinberg,
\newblock Nucl. Phys. {\bf B363}, 3 (1991).

\bibitem{Weinberg:1992yk}
S.~Weinberg,
\newblock Phys. Lett. {\bf B295}, 114 (1992), hep-ph/9209257.

\bibitem{Ordonez:1996rz}
C.~Ordonez, L.~Ray, and U.~van Kolck,
\newblock Phys. Rev. {\bf C53}, 2086 (1996), hep-ph/9511380.

\bibitem{Kaplan:1998tg}
D.~B. Kaplan, M.~J. Savage, and M.~B. Wise,
\newblock Phys. Lett. {\bf B424}, 390 (1998), nucl-th/9801034.

\bibitem{Kaplan:1998we}
D.~B. Kaplan, M.~J. Savage, and M.~B. Wise,
\newblock Nucl. Phys. {\bf B534}, 329 (1998), nucl-th/9802075.

\bibitem{Epelbaum:1998na}
E.~Epelbaum, W.~Glockle, A.~Kruger, and U.-G. Meissner,
\newblock Nucl. Phys. {\bf A645}, 413 (1999), nucl-th/9809084.

\bibitem{Epelbaum:1998hg}
E.~Epelbaum, W.~Glockle, and U.-G. Meissner,
\newblock Phys. Lett. {\bf B439}, 1 (1998), nucl-th/9804005.

\bibitem{Epelbaum:1998ka}
E.~Epelbaum, W.~Glockle, and U.-G. Meissner,
\newblock Nucl. Phys. {\bf A637}, 107 (1998), nucl-th/9801064.

\bibitem{Epelbaum:2002vt}
E.~Epelbaum {\em et~al.},
\newblock Phys. Rev. {\bf C66}, 064001 (2002), nucl-th/0208023.

\bibitem{Lee:2003a}
D.~Lee and P.~Maris,
\newblock Phys. Rev. {\bf D67}, 076002 (2003).

\bibitem{Ipsen:2003}
I.~C. Ipsen and D.~Lee,
\newblock Determinant approximations,
\newblock Submitted for publication, 2003.

\bibitem{Borasoy:2003}
B.~Borasoy, D.~Lee, and T.~Schaefer,
\newblock Work in progress, 2003.

\bibitem{Soper:1978dp}
D.~E. Soper,
\newblock Phys. Rev. {\bf D18}, 4590 (1978).

\bibitem{Rothe:1997kp}
H.~J. Rothe,
\newblock World Sci. Lect. Notes Phys. {\bf 59}, 1 (1997).

\bibitem{Hirsch:1983}
J.~E. Hirsch,
\newblock Phys. Rev. {\bf B28}, 4059 (1983).

\bibitem{Goldberger:1958vp}
M.~L. Goldberger and S.~B. Treiman,
\newblock Phys. Rev. {\bf 111}, 354 (1958).

\bibitem{Elliott:2001hn}
ISiS, J.~B. Elliott {\em et~al.},
\newblock Phys. Rev. Lett. {\bf 88}, 042701 (2002), nucl-ex/0104013.

\bibitem{Ma:2003dc}
Y.~G. Ma {\em et~al.},
\newblock (2003), nucl-ex/0303016.

\bibitem{Karnaukhov:2003vp}
V.~A. Karnaukhov {\em et~al.},
\newblock Phys. Rev. {\bf C67}, 011601 (2003), nucl-ex/0302006.

\bibitem{Chandrasekharan:2003wy}
S.~Chandrasekharan, M.~Pepe, F.~D. Steffen, and U.~J. Wiese,
\newblock (2003), hep-lat/0306020.

\end{thebibliography}

\end{document}